\documentclass{article}

\usepackage{arxiv}

\usepackage[utf8]{inputenc} % allow utf-8 input
\usepackage[T1]{fontenc}    % use 8-bit T1 fonts
\usepackage{hyperref}       % hyperlinks
\usepackage{url}            % simple URL typesetting
\usepackage{booktabs}       % professional-quality tables
\usepackage{amsfonts}       % blackboard math symbols
\usepackage{nicefrac}       % compact symbols for 1/2, etc.
\usepackage{microtype}      % microtypography
\usepackage{lipsum}
\usepackage{cite}
\usepackage{amsmath}
\usepackage{float}
\usepackage{graphicx}

\title{The gray body approximation for radiative heat transfer in evacuated tube solar collectors: Effects of envelope infrared transparency}

\author{
  Mark A. George\thanks{Corresponding authors }\\
  University of Sydney\\
  Sydney, New South Wales, 2006 Australia \\
  \texttt{mgeo2280@uni.sydney.edu.au} \\
   \And
 Noboru Takamure \\
  School of Physics\\
  University of Sydney\\
  Sydney, New South Wales, 2006 Australia  \\
  \And
    David R. McKenzie$^*$ \\
  School of Physics\\
  University of Sydney\\
  Sydney, New South Wales, 2006 Australia  \\
  \texttt{david.mckenzie@sydney.edu.au} \\
}

\begin{document}
\maketitle

\begin{abstract}
A theoretical and experimental analysis is carried out of radiative heat transfer in the coaxial geometry of evacuated tube solar collectors. The gray body approximation implicit in the use of an effective emissivity does not strictly apply to evacuated tube solar collectors due to selective absorber coating and partially transmitting outer glass in the thermal infrared, especially when constructed from borosilicate. We develop analytic expressions for the heat transfer through the outer envelope and show the equations no longer follow a simple form where an effective emissivity for the system can be defined. To test all approximations in practice, an experiment is performed using an evacuated solar collector manufactured in the 1980s by the Nitto Kohki company in Japan using the effective emissivity approximation to determine the typical heat transfer characteristics using net radiative heat flows in both directions. This method enabled a good fit to temperature–time data for cooling and heating of the inner tube for temperatures between 10 and 85 $^\circ$C. The results confirm that the effective emittance method can be used in situations with typical glass wall thickness and temperatures. At temperatures greater 100$^\circ$C, the spectral distribution of the emitted radiation falls significantly within the transmitting region of the outer glass and can no longer be neglected. The work has verified the stability of the vacuum in this type of collector as the tube still functions well, maintaining a low emissivity and good vacuum after approximately 40 years in storage conditions.
\end{abstract}

% keywords can be removed

%%%%%%%%%%%%%%%%%%%%%%%%%%%%%%%%%%%%%%%%%%%%% Introduction %%%%%%%%%%%%%%%%%%%%%%%%%%%%%%%%%%%%%%%%%%%%%%%%%
\section{Introduction}
The evacuated tube solar collector is now common around the world as a source of heat for domestic \cite{Morrison1984} and industrial \cite{Grass2004} purposes. The design originating from the University of Sydney \cite{Collins1986} consists of an all-glass single-ended coaxial design with a sputtered selective surface composed of a copper base layer and a stainless steel-carbon absorbing layer on an inner tube and a vacuum seal on an outer envelope. The tubes are fused together at one end, where the interior volume of the inner tube is open to the air and both tubes have a hemispherical termination at the other end. This design was manufactured under licence to the University of Sydney in the 1980s by the Nitto Kohki company and subsequently by the Shiroki Company, both of Japan. The construction material is borosilicate glass with an outer envelope of wall thickness in the range of 1–3 mm. In operation, the inner tube contains a heat transfer medium such as water,\cite{Zhiqiang1984} air, or other fluid, that may fill the inner tube or be contained in metal pipes thermally connected to the interior surface of the inner tube\cite{Morrison1984, Mcphedran1983}. One of the applications of the evacuated tube collector is for producing high-temperature fluids such as air \cite{Wang2014} and superheated water. Operation at high temperatures increases the amount of heat transmitted through the borosilicate glass because of its partial transparency windows in the thermal infrared. It is possible, depending on the thickness of the outer glass envelope, that this partial transparency effect should be taken into account in heat loss calculations. 

The concentric tube geometry has also become of interest in the encapsulation of high performance solar cells, such as perovskite solar cells \cite{Park2015, Granados2020, Shi2020} that are vulnerable to degradation by environmental exposure to moist air. Encapsulation of the cells in a tubular geometry under controlled atmosphere would enable the cells to be located on the outer surface of the inner tube, in the same way as the selective coating is applied to the conventional evacuated tube collector. However, the heat control problem in this application would be to keep the solar cell cool to maintain its performance and reduce thermal degradation effects, since the performance of such cells decreases with increasing operating temperature. The radiative transfer problem is related to that in the conventional tube collector but becomes one of the maximizing rather than minimizing the radiative transfer from the surface of the inner tube, onto and through the outer envelope.

A considerable amount of research has been done on evacuated solar collectors to evaluate their heat extraction efficiency in the context of their intended application using experimental and numerical methods \cite{Ma2010, Arora2011, Harding1985, Zambolin2010, Morrison2004} However, in the analysis of radiative heat transfer between the inner and outer tubes, it is normally assumed that an effective emittance can be used to describe radiative transfers between the inner and outer envelopes, which takes into account the individual emittances of the two surfaces. There are two areas where the conventional approach to calculating radiative heat transfers in the evacuated tubular collector geometry could be deficient. First, the gray body approximation, which allows the radiative heat transfer to be written as a Stefan Law expression consisting of the difference of the fourth powers of the inner and outer tube temperatures, is subject to challenge. The surface coating on the inner tube has an absorbance and, therefore, an emittance that are spectrally dependent and so the inner tube is far from a gray body, having a strong dependence of the emittance on wavelength in the region near 1–5 $\mu$m where the emittance changes from very high to relatively low. The second area where existing treatments are potentially deficient is the known transparency of the outer envelope to the thermal infrared, neglected in previous studies. The coaxial geometry has also been used to determine a value of the total hemispherical emittance of sputtered copper films \cite{Window1981}. When used for this kind of fundamental measurement, careful account of the outer envelope contributions to the radiative heat transfer must be taken into consideration. 

While there has been a study of the in-service performance of these tubes, specifically looking at the degradation of the vacuum due to atmospheric permeation and outgassing of selective coating \cite{Chow1985}, this was an accelerated aging study rather than a real-time study. 

In this work, we develop equations based on the gray body approximation while allowing transparency of the outer envelope. We estimate the effect of transparency by estimating the error that would arise from the conventional assumption of an effective emittance. Then, we test the combined effect of both areas of approximation in a practical situation by undertaking an experimental study of the heating and cooling curves using an evacuated collector manufactured by the Nitto Kohki company in the 1980s. The experiment was performed on a tube selected randomly from a batch that has been in long-term storage since it was manufactured. The experiment has an additional purpose to verify the long-term durability of the evacuated tube concept in terms of the stability of the selective absorber’s low emittance and of the ability of the all-glass envelope to maintain a sufficient level of vacuum to avoid conductive heat transfers. The experiment was performed by allowing a fluid within the tube to heat up or cool down and then measuring the temperature–time relationship of the inner and outer surfaces.

%%%%%%%%%%%%%%%%%%%%%%%%%%%%%%%%%%%%%%%%%%%%%%%%% Theory %%%%%%%%%%%%%%%%%%%%%%%%%%%%%%%%%%%%%%%%%%%%%%%%%%%
\section{Theory}
\subsection{Background}
The heat transfers in an evacuated tube collector take place between two cylindrical coaxial bodies, the inner one at temperature $T_1$ and the outer one at temperature $T_2$. It is assumed that the cylinders have infinite length along their axis. Only the case where the two coaxial bodies are non-transmitting to thermal infrared radiation has been considered for applications in solar collectors and an analogous approximation has been made for the parallel plate geometry in vacuum glazings \cite{Ma2010, Window1981}. The radiative transfer between two coupled, infinitely extended planes with opaque surfaces has been discussed in textbooks such as those of Holman \cite{Holman1997} and Howell et al. \cite{Howell2010} Applying the gray body approximation in which the total hemispherical absorptivity is assumed not to depend on the spectral distribution of the incoming radiation, the rate of the radiative heat flux between the two surfaces is given by the Stefan radiation law in the following form:
\begin{align}
    q_1 = \varepsilon_{eff} \sigma (T_1^4 - T_2^4), \label{eq:1}
\end{align}
where $T_1$ and $T_2$ are the temperatures of each surface and $\sigma$ is the Stefan Boltzmann constant. This equation relies on the assumption that the radiating bodies are ``fully coupled" meaning that emitted photons from one surface are always intercepted by the other, and the effective emissivity is given by:
\begin{align}
    \varepsilon_{eff} = \left( \frac{1}{\varepsilon_1 (T_1)} + \frac{1}{\varepsilon_2 (T_2)}-1 \right)^{-1}, \label{eq:2}
\end{align}
where $\varepsilon_1 (T_1)$ and $\varepsilon_2 (T_2)$ are the total hemispherical emissivities of each body at their respective temperatures. In the case where the two bodies are infinite coaxial cylinders, the bodies are not fully coupled and some of the radiation emitted by the outer body does not intercept the inner. The effective emissivity in the coaxial geometry then becomes:
\begin{align}
    \varepsilon_{eff} = \left( \frac{1}{\varepsilon_1 (T_1)} + \frac{R_1}{R_2} \left( \frac{1}{\varepsilon_2(T_2)} - 1 \right) \right)^{-1}. \label{eq:3}
\end{align}
Where $R_1$ and $R_2$ are the radii of the inner and outer coaxial tubes, respectively. Both formulas are derived by considering all possible reflections of emitted photons and summing the contributions of all reflections to infinity. This expression has been used before in the analysis of evacuated solar collectors \cite{Ma2010}.

Equation \ref{eq:3} applies only to radiative transfers between nontransmitting surfaces. In the case of an evacuated collector (or a vacuum glazing), the outer surface may be partially transmitting, and consequently, the concept of an effective emittance is no longer applicable, and the heat transfer follows a different form as shown. Here, we derive the equations for the two-body coaxial geometry where the outer cylinder is partly absorbing and partly transmitting for the thermal radiation, i.e., an ``infrared window''. We show that the resulting expression is different to that given in Eqs. \ref{eq:1} and \ref{eq:3} and involves the transmittance of the outer envelope, which has direct application to evacuated solar collectors.

\subsection{Effect of transmission through the outer envelope}
\begin{figure}[H]
    \centering
    \includegraphics[width=0.75\linewidth]{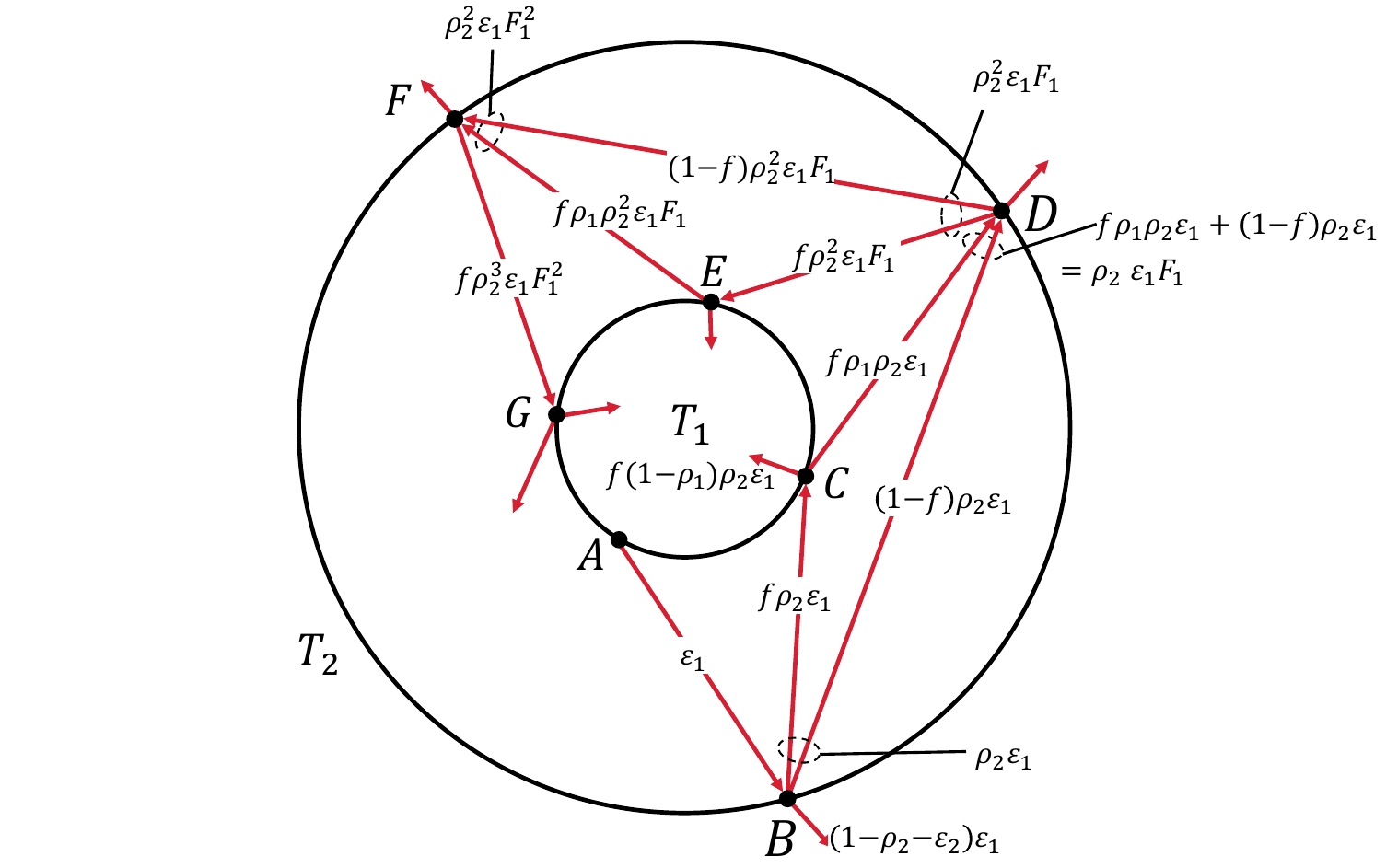}
    \caption{Possible paths of a single ray of radiation initially emitted from the surface of the inner tube at point $A$. Upon reflection from the outer tube at $B$, the ray can either coincide again with the outer tube at $D$ or undergo another reflection at $C$ to later coincide with the outer tube.}
    \label{fig:Fig1}
\end{figure}

\begin{figure}[H]
    \centering
    \includegraphics[width=0.75\linewidth]{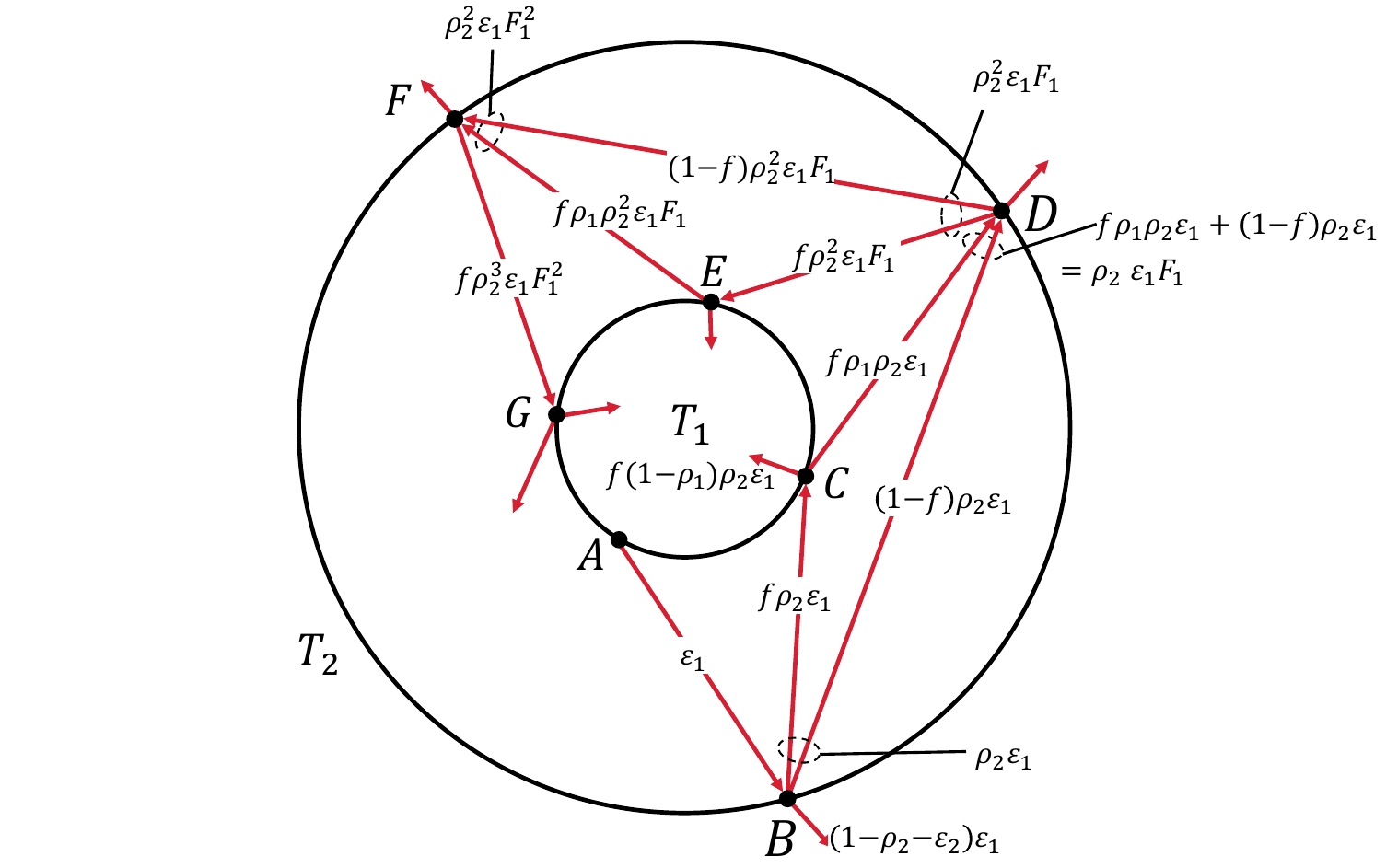}
    \caption{Possible paths of a single ray originating from the outer tube at point $A$. As in Fig. \ref{fig:Fig1}, the ray can take various paths within the tube and will undergo, in principle, an infinite number of reflections.}
    \label{fig:Fig2}
\end{figure}

The expression for $Q_r$, the net radiation out of the inner tube, is now derived while accounting for the transmissive outer tube. The temperature dependence of all surface optical properties is neglected and is assumed to be constant within the working temperature range of the collector. The spectrum of the radiation emitted from the inner tube is not black body, since the surface is spectrally selective. The total reflectivity of the outer tube for this radiation depends on the spectral properties of the inner surface and will also be modified after reflection from the outer surface, if the outer surface is also spectrally selective (see Figs. \ref{fig:Fig1} and \ref{fig:Fig2}). The spectral content of the radiation evolves with the number of reflections from the outer and inner surfaces and the problem becomes a difficult one to solve exactly. The \textit{gray body approximation} is now invoked, meaning that the evolving nature of the reflected radiation between the tubes is ignored and the total hemispherical reflectivities of both surfaces are considered as constants. For the inner tube (body 1), the total hemispherical emissivity is equated to the total hemispherical absorptivity since it is considered non-transmissive, and using an angular and spectrally integrated form of Kirchoff’s law, we write $\varepsilon_1 = \alpha_1 = 1 - \rho_1$. For the outer tube (body 2), transparency is allowed, and as with emissivity, a spectrally averaged transmittance is considered. The reflectivity for the outer tube is given by $\rho_2 = 1 - \tau_2 - \varepsilon_2$, where $\tau_2$ is the total hemispherical transmissivity of the outer glass. These properties and, in particular, the transmittance can change with temperature due to their spectral dependence and must be accounted for accordingly based on the operating temperature of the tube. Furthermore, these properties can effectively be seen to depend on the incident radiation on the outer envelope from the inner tube—if the incident radiation lies spectrally within the transmitting region of the glass envelope, the transmissivity is high. We define $f = R_1/R_2$, where $R_1$ and $R_2$ are the radii of the inner and outer tubes, respectively. The use of the gray body assumption and the integrated form of Kirchoff’s law are justified ultimately by the agreement with observation. 

To determine $Q_r$, we sum contributions from rays that are emitted from either the inner tube or the outer tube and then undergo an infinite number of reflections. First, considering a ray with intensity $\sigma A_1 T_1^4$ emitted from the inner tube (Fig. $\ref{fig:Fig1}$) at point $A$, its intensity will gain a factor of $\varepsilon_1$ due to the emission and undergo a reflection at $B$. At each reflection from the outer tube, some of the radiation is absorbed or transmitted, the rest is reflected, gaining a factor of $\rho_2$. This reflection may either intercept the inner tube with fraction f or the outer tube again with fraction $(1-f)$. Both these rays will impinge upon the outer tube again upon reflections with decreased intensity. Accounting for an infinite number of reflections, the radiation re-absorbed by the inner tube after being emitted from it is given by summing the contributions,
\begin{align}
    Q_{r_1} &= -\alpha A_1 \sigma (f \rho_2 \varepsilon_1 + f \rho_2^2 \varepsilon_1 F_1 + f \rho_2^3 \varepsilon_1 F_1^2 + ...)T_1^4 \nonumber \\
    &= -\frac{f \rho_2 \varepsilon_1^2}{1 - \rho_2 F_1} A_1 \sigma T_1^4, \label{eq:4}
\end{align}
where a geometric series and the relation $\varepsilon_1 = \alpha_1$ is used to simplify the expression along with $F_1 = 1 - f + f \rho_1$.

The contribution due to radiation initially emitted from the outer tube is now examined. Referring to Fig. \ref{fig:Fig2}, a ray with intensity of $\sigma A_2 T_2^4$ is emitted from point $A$ and will undergo the same pattern as previously discussed. Hence, the total absorbed power into the inner tube by rays emitted from the outer tube is given by
\begin{align}
    Q_{r_2} &= - \alpha_1 A_2 \sigma( f \varepsilon_2 + f \rho_2 \varepsilon_2 F_1 + f r_2^2 F_1^2 \varepsilon_2 + ... )T_2^4 \nonumber \\
    &= -\frac{\varepsilon_1 \varepsilon_2 f}{1 - \rho_2 F_1} A_2 \sigma T_2^4, \label{eq:5}
\end{align}
where again a geometric series and $\varepsilon_1 = \alpha_1$ are used. Accounting for radiation emitted by the tube initially given by $\varepsilon_1 A_1 \sigma T_1^4$ , summing the contributions given by Eqs. \ref{eq:4} and $\ref{eq:5}$, the result is
\begin{align}
    Q_r = A_1 \sigma \left( \frac{\varepsilon_1(1 - \rho_2 F_1) - f \rho_2 \varepsilon_1^2 }{1 - \rho_2 F_1} \right) T_1^4 - A_2 \sigma \left( \frac{\varepsilon_1 \varepsilon_2}{1 - \rho_2 F_1} \right) T_2^4.\label{eq:6}
\end{align}
Note that this expression cannot be reduced to the form of Eq. \ref{eq:1}. In the case that $\tau_2 = 1 - \varepsilon_2 - \rho_2 = 0$, Eq. \ref{eq:3} is obtained. If further $f = 1$, Eq. \ref{eq:2} is obtained. Hence, Eq. \ref{eq:6} matches known limiting cases. Since the average emittance of glass is greater than 0.85 at 300 K \cite{Jones2019} and the reflectance of glass is less than 0.1 for most wavelengths, \cite{El-Zaiat2013} we can approximate the average transmissivity of the outer layers as zero. The validity of this approximation that greatly simplifies the equations and allows the use of the effective emissivity to determine the heat transfers between tubes as assumed previously is further discussed in Sec. \ref{sec:discussion}.

\subsection{Time dependence of temperature including conductive transfers}
\begin{figure}[H]
    \centering
    \includegraphics[width=0.75\linewidth]{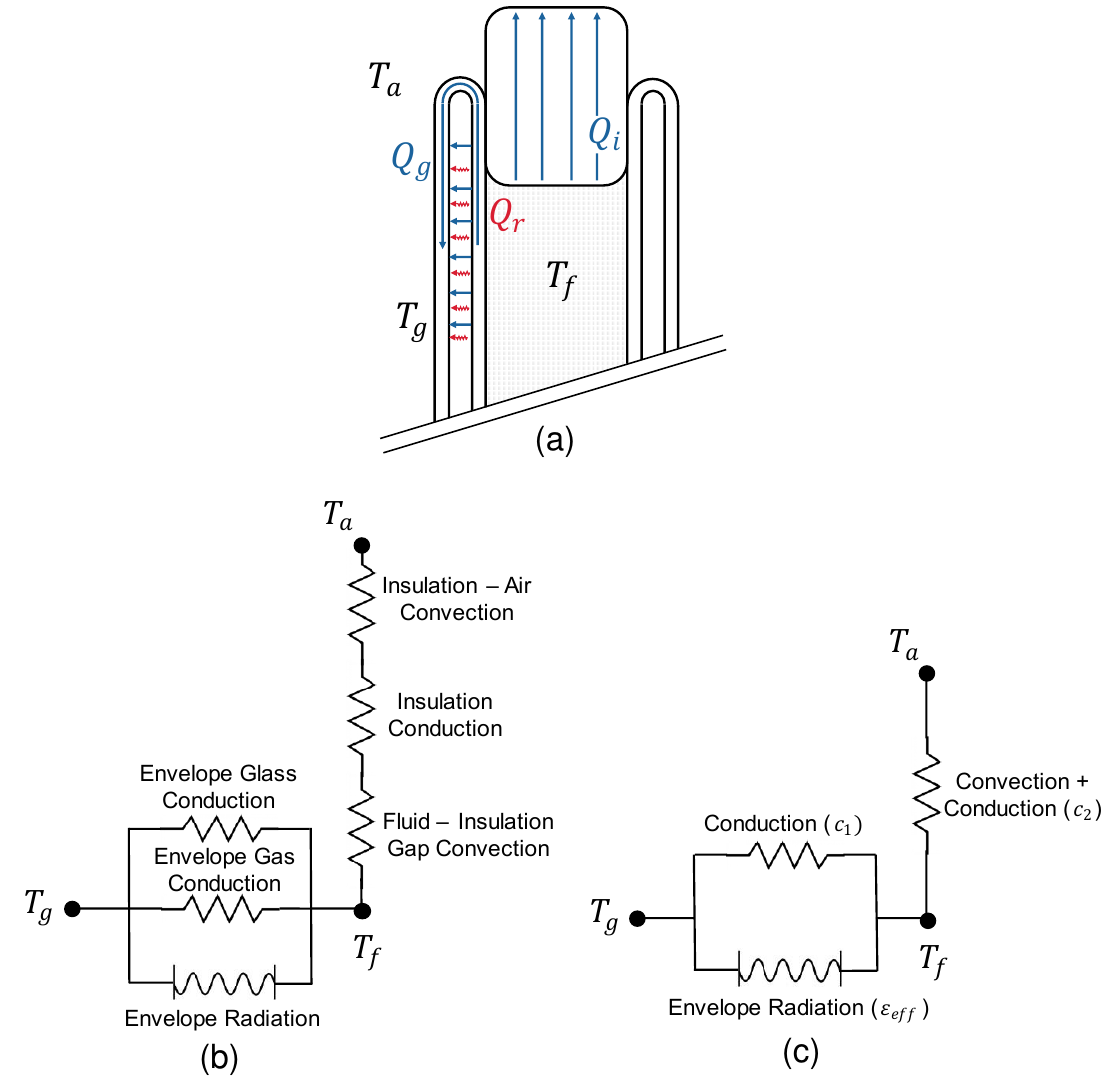}
    \caption{Heat escapes the water in three main ways (a): Radiation through the vacuum envelope $(Q_r)$, conduction through the insulating cap $(Q_i)$, and conduction through the glass around the envelope and through residual gas in the vacuum $(Q_g)$. $T_a$ is the temperature of the ambient air, $T_f$ is the temperature of the fluid within the tube, and $T_g$ is the temperature of the outer glass layer. The inside of the tube is filled with a fluid (water or ethanol), as indicated by the shaded region. Panel (b) is a representative thermal circuit of the system, and panel (c) is a simplified thermal circuit used to create the thermal model.}
    \label{fig:Fig3}
\end{figure}

If the tube is filled with a fixed mass of a fluid, the effective emissivity is determined by considering the total heat lost or gained by the fluid. The lumped capacitance model for the fluid is assumed, where the fluid is treated as a single mass of uniform temperature. This assumption is valid given that the thermal resistance to conduction and convection within the fluid is much less than resistance to radiation at the boundary of the fluid. \cite{Incropera2007} The validity of the lumped capacitance model is verified through the calculation of the Biot number for radiation in Sec. \ref{sec:results}. The heat energy per unit time lost or gained by the fluid is given by
\begin{align}
    \frac{\partial E}{\partial t} = m_f c_p(T_f) \frac{\partial T_f}{\partial t}, \label{eq:7}
\end{align}
where $n_f$ is the mass of the contained fluid and $c_p(T_f)$ is the specific heat capacity of the fluid at constant pressure (water and ethanol for this experiment) as a function of temperature. It is assumed that the change in volume of the fluid throughout the experiment is negligible. The total energy lost by the fluid is the energy radiated $Q_r$, the energy conducted through the insulating cap, $Q_i$, and the energy conducted through the glass envelope $Q_g$ . These heat transfers are depicted in Fig. \ref{fig:Fig3}(a). Figure \ref{fig:Fig3}(b) shows a thermal circuit model of the tube system. The conducted energy between the fluid and glass envelope can come from any residual gas within the tube, but also around the glass envelope, which can be represented as a parallel resistance. Heat transfers between the fluid and the air through the insulating cap go through various paths in series. Between the fluid and the base of the insulating gap, there are convective transfers through the small air gap; furthermore, there is natural convection occurring with the ambient air and the top of the insulating cap. The thermal circuit can be simplified to the one shown in Fig. \ref{fig:Fig3}(c). Taking the conduction and convection to have a linear dependence on the temperature difference, an energy balance on the fluid gives the differential equation,
\begin{align}
    m_f c_p (T_f) \frac{\partial T_f}{\partial t} = \varepsilon_{eff} A_1 \sigma (T_g^4 - T_f^4) + c_1 (T_g - T_f) + c_2(T_a - T_f). \label{eq:8}
\end{align}
The coefficients $c_1$ and $c_2$ that incorporate convection depend on the problem geometry, boundary conditions, and fluid properties; and they are not necessarily constant with temperature \cite{Incropera2007}. The approximation that they are constant is made here since these parameters vary only a small amount over the duration of the experiment. Equation \ref{eq:8} has three free parameters; $\varepsilon_{eff}$, $c_1$, and $c_2$. The effective emissivity will be the same across all experiments since it depends on tube geometry and surface properties only. The parameter $c_1$ is the heat transfer coefficient between the fluid and the outer glass. This parameter depends on the thermal conductivity of the glass tube, the geometry of the glass, and the residual gas in the envelope; hence, it takes the same value across all experiments. It is expected that $c_1$ will be very close to zero for two reasons. First, the borosilicate glass used to construct the tube has a low thermal conductivity, with a small cross sectional area (1.5 mm glass thickness) and long conduction path (450 mm tube length). Conduction through glass will have large thermal resistance. 

\begin{figure}[H]
    \centering
    \includegraphics[width=0.45\linewidth]{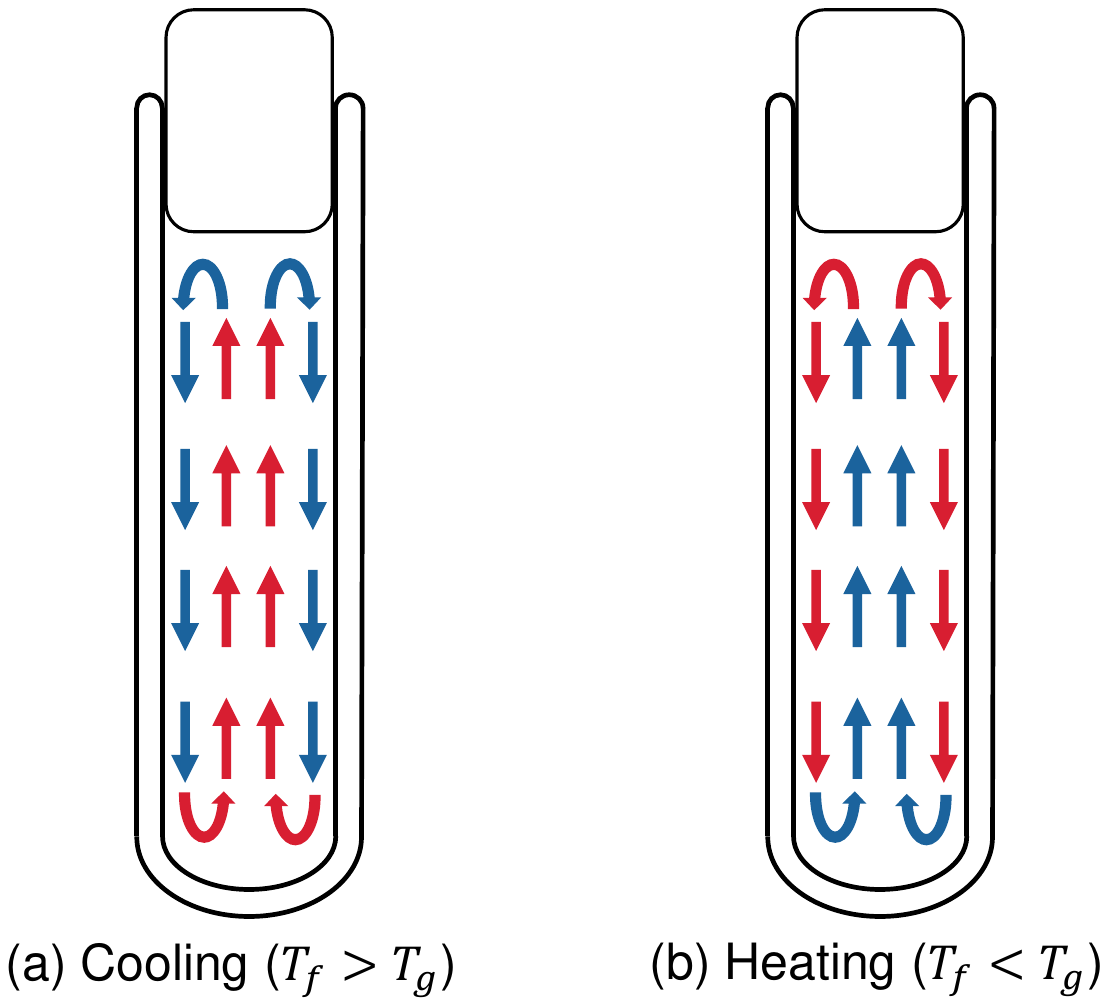}
    \caption{Buoyancy-driven convection loops result in stratification of the water within the tube. These loops depend on whether the fluid is being cooled (a) or heated (b) \cite{Lin1999, Wenxian2001, Evans1968}.}
    \label{fig:Fig4}
\end{figure}

Second, if there is little residual gas within the vacuum envelope,  convective transfers between the envelope will be negligible. The parameter $c_2$ relates heat transfers between the ambient air and the fluid. As shown in the thermal circuit [Fig. \ref{fig:Fig3}(b)], there is convection in the small air gap between the insulation and the working fluid, but also to the ambient air. The coefficients that describe such convective transfers are dependent on geometry, fluid properties, and the direction of the heat flow with respect to gravity \cite{Incropera2007}. For the experiments done here, the fluid temperature can be below ambient (as is the case with heating the fluid) or above ambient (when cooling the fluid); hence, the convection coefficient to the air will vary. Furthermore, the fluid within the tube will become thermally stratified upon cooling and heating. Depending on whether the fluid is being cooled or heated, different convective loops and temperature distributions will form. These temperature distributions will be a function of the heating rates, heating direction, and fluid properties such as viscosity, density, and thermal conductivity \cite{Lin1999, Wenxian2001, Evans1968} A representative diagram of the convective patterns that may form in the tube is shown in Fig. \ref{fig:Fig4}. Consequently, the convective transfer resistances in Fig. \ref{fig:Fig3}(b) will depend on the fluid used and the heating direction. As a result, the coefficient $c_2$ should be different for each case where the direction of the heat flow of the working fluid within the tube is changed.
 
 Equation \ref{eq:8} is fitted to the experimental temperature–time cooling curves using the least squares method to determine the optimal parameters, with $\varepsilon_{eff}$ and $c_1$ being the same across all experiments and $c_2$ being allowed to vary for each case of water or ethanol filling, heating, and cooling (four cases in total). The coefficients in this fitting process are unconstrained. The cooling curves (temperature vs time) were used to calculate the heat radiated between the selective coating on the inside of the vacuum envelope and glass on the outside of the envelope. This allows a measurement of the effective emissivity of the system. As well as the heat radiated in the vacuum envelope, there are conductive heat transfers from the water. The main paths of conduction that cannot be neglected are those through the insulating cap and between the layers of the glass envelope due to possible residual gas.

%%%%%%%%%%%%%%%%%%%%%%%%%%%%%%%%%%%%%%%%%% Experimental Methods %%%%%%%%%%%%%%%%%%%%%%%%%%%%%%%%%%%%%%%%%%%%
\section{Experimental Methods}
\begin{figure}[H]
    \centering
    \includegraphics[width=0.65\linewidth]{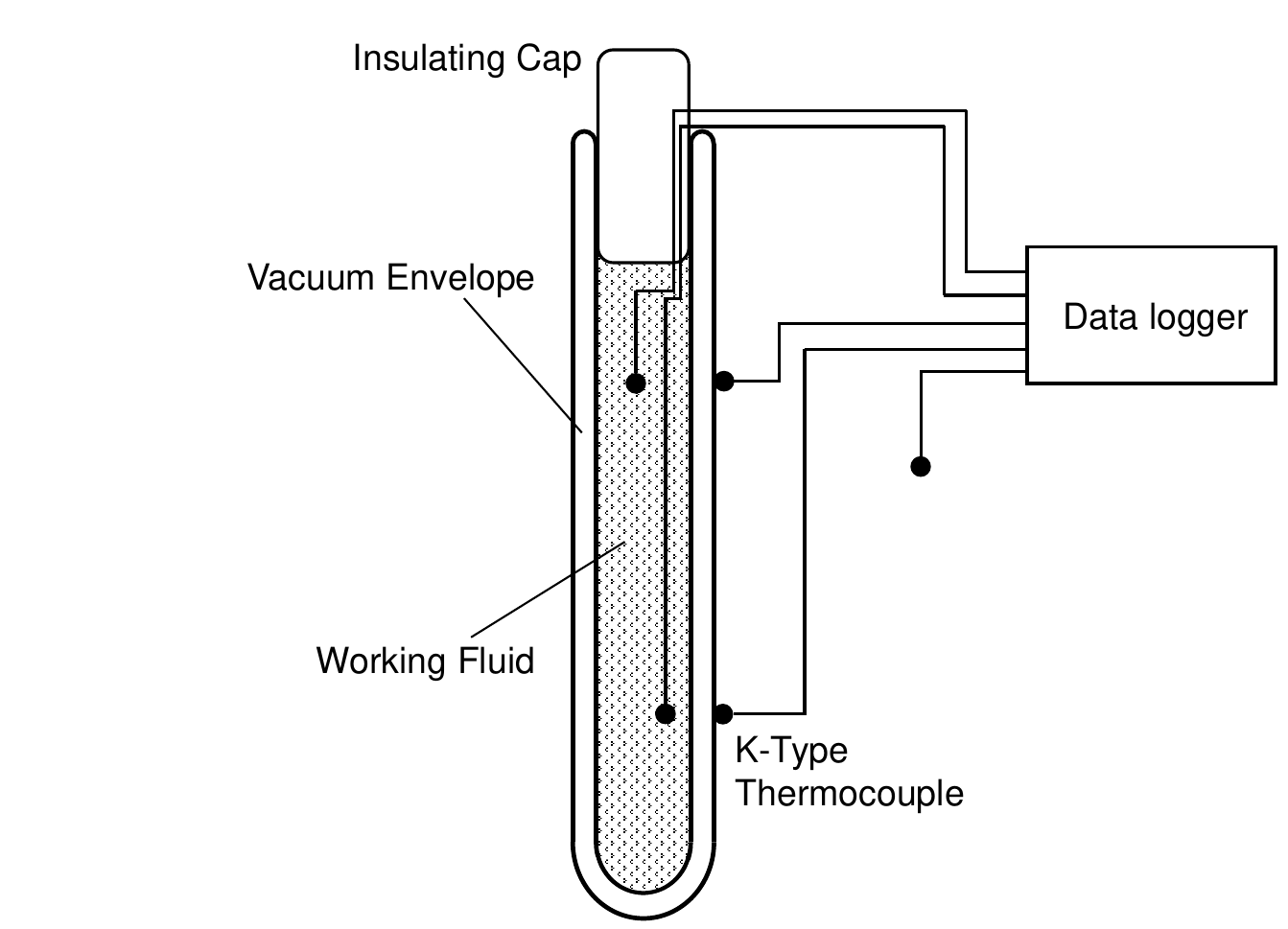}
    \caption{Experimental setup used to obtain cooling curves for the evacuated tube solar collector.}
    \label{fig:Fig5}
\end{figure}

The evacuated tube was set up vertically away from direct sunlight, other strong light sources, or heat sources and was placed close to the ground. A schematic of the experiment is shown in Fig. \ref{fig:Fig5}.

The temperatures being measured are the temperature of the fluid at two locations, and the temperature of the outside glass surface at two different locations. The average of the two is used in the calculation. It was found that throughout the life of the experiment, the temperature difference between both thermocouples did not differ by more than 0.5$^\circ$C. It is assumed that these measured temperatures are sufficiently close to the temperature of the actual radiating surfaces.

Two versions of the experiment were done; one where the net heat radiation flows from the working fluid to the outside ambient air, the other where the net heat radiation flows in the opposite direction from the ambient air to the working fluid. Water and ethanol were both used as working fluids for additional confirmation and for estimating errors in the experiment.

When performing the experiment with the heated working fluid, the tube was filled with boiling water to preheat the tube. Preheating the tube means less heat in the working fluid will be lost to the mass of the glass, and the cooling curves will be more representative of a radiation curve. After preheating, the tube was emptied and filled with the working fluid at temperature and then sealed with an insulating cap made from EVA foam. Two K-Type thermocouples were placed within the tube to measure the internal water temperature, and two were placed on the outside surface of the glass to measure the glass temperature. A single K-Type thermocouple was also used to measure the ambient temperature of the air surrounding the tube. Measurements were logged every 5 s, using MAX31855 temperature sensors, which provide an accuracy of up to 0.25$^\circ$C. The experiment was run over at least a 72 h period to ensure that thermal equilibrium has been reached with the ambient temperature.

For the net heat flow in the opposite direction for the second experiment, a cylindrical copper tube was placed around the collector and sealed from both ends using densely packed open cell foam. The copper tube was then heated with a ribbon heater and a NOVUS N322 feedback controller. Thermocouple placement was the same. Cool fluid was poured into the tube, and after being left to reach equilibrium, it was sealed with the same foam insulating cap. Surrounding the collector in a copper tube may result in different background radiation entering the tube through the outer envelope. This effect was deemed negligible after the cooling experiment was conducted with the copper tube around it, and the same results were obtained.

The cooling experiments were done from a temperature just below the boiling point of the fluid to ambient temperature of the room. Heated fluid was placed inside the tube and allowed to cool. For the heating experiments, the tube was filled with a chilled fluid above the freezing temperature and allowed to heat to a temperature below its boiling point. That is, the copper heating tube was set to heat the outer glass layer to a temperature a small amount below the boiling point of the fluid. It is important that no phase change of the fluid occurs during the experiment.

%%%%%%%%%%%%%%%%%%%%%%%%%%%%%%%%%%%%%%%%%%%%%%% Results %%%%%%%%%%%%%%%%%%%%%%%%%%%%%%%%%%%%%%%%%%%%%%%%%%%%
\section{Results} \label{sec:results}
When the experimental measurements were fit to theory, the initial sections of data corresponding to the start of the experiment and the end section corresponding to the system reaching equilibrium were removed. This avoids errors that may arise due to initial turbulence in the fluid when preparing the experiment, and environmental noise that may arise when the system is in thermal equilibrium. In MATLAB, the fit was performed using the \textit{lsqcurvefit} function. The \textit{nlparci} function was used to calculate 95\% confidence intervals in the fitted parameters, which is taken as uncertainty. The results are shown in Table \ref{tab:1}.
\raggedbottom
\begin{table}[H]
 \caption{Fitted parameters in Eq. \ref{eq:8} to the measured data using least squares. The uncertainties are the 95\% confidence intervals on the parameters for the least squares fit}
 \label{tab:1}
  \centering
  \begin{tabular}{lllll}
    \toprule
         & Water - cooling     & Water - heating & Ethanol - cooling & Ethanol - heating \\
    \midrule
    $\varepsilon_{eff}$ & \multicolumn{4}{c}{$(7.11 \pm 0.03)\times 10^{-2}$}  \\
    $c_1$ (W/K)     &   \multicolumn{4}{c}{$(0.00 \pm 0.09) \times 10^{-3}$}  \\
    $c_2$ (W/K)     &  $(5.27 \pm 0.01)\times 10^{-3}$  & $(1.29 \pm 0.01)\times 10^{-3}$ & $(5.30 \pm 0.02)\times 10^{-3}$ & $(3.77 \pm 0.02)\times 10^{-3}$ \\
    \bottomrule
  \end{tabular}
\end{table}

\begin{figure}[H]
    \centering
    \includegraphics[width=0.75\linewidth]{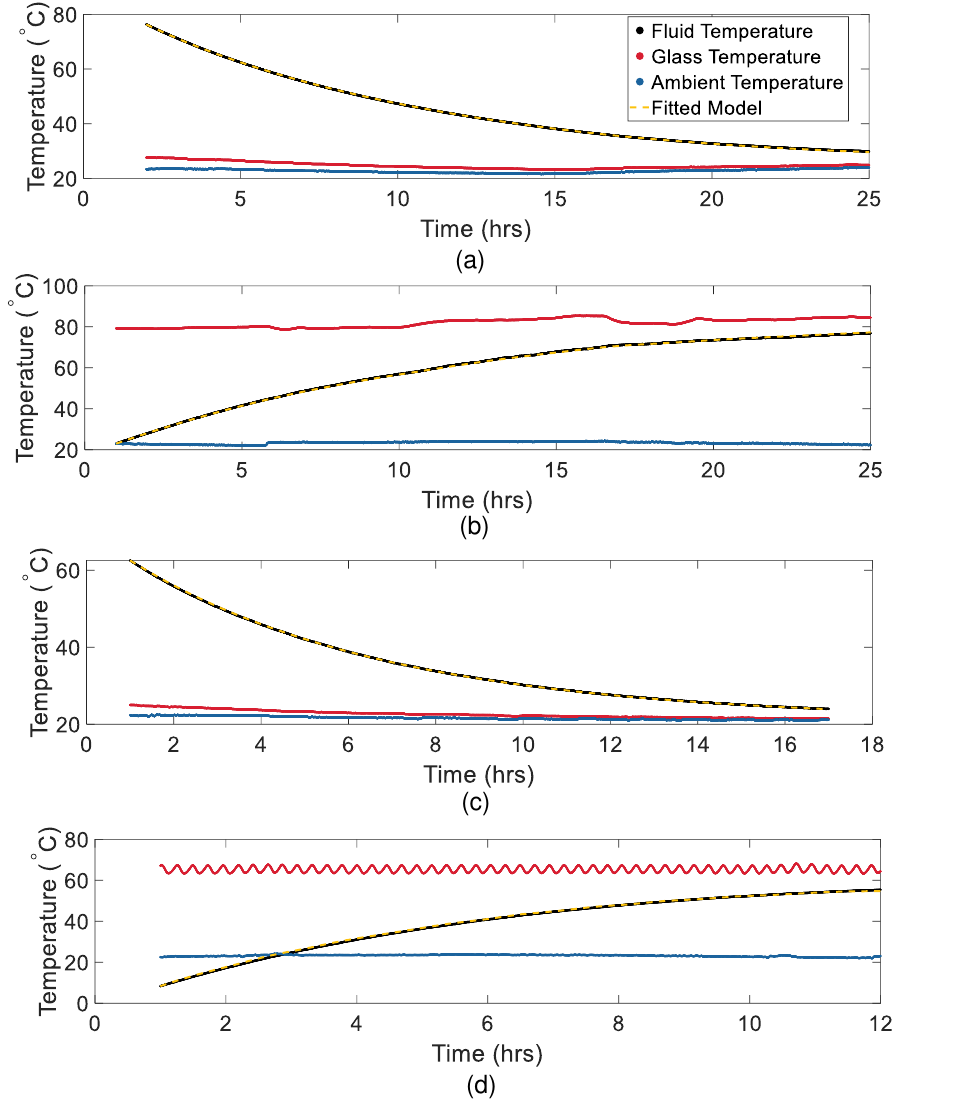}
    \caption{Measured temperatures compared with predicted temperatures according to the fitted model given by Eq. \ref{eq:8} for the fluid temperature using parameters in Table \ref{tab:1}. (a) and (b) are the cooling and heating experiments, respectively, with water; (c) and (d) are the cooling and heating experiments, respectively, with ethanol.}
    \label{fig:Fig6}
\end{figure}

Knowledge of the parameters in Table \ref{tab:1} allows one to determine the transient behavior of the temperature of the fluid given the initial temperature and the boundary condition, in this case, the boundary condition is the outer glass temperature. The results for solving Eq. \ref{eq:8} numerically with the fitted parameters for each experiment case are shown in Fig. \ref{fig:Fig6}. In the heating experiments [Figs. \ref{fig:Fig6}(b) and \ref{fig:Fig6}(d)], the glass temperature exhibits oscillations, particularly in the ethanol experiment; this is a consequence of the on-off feedback controller used to maintain the outer glass envelope at the elevated target temperature. The temperature will oscillate as the input is switched on and off to maintain the temperature within the hysteresis band.

The Root Mean Squared (RMS) deviation of the fitted curve from the measured data from curves (a), (b), (c), and (d) are 0.0997, 0.243, 0.0789, and 0.398, respectively. Since Eq. \ref{eq:8} is linear in the parameters $\varepsilon_{eff}$ , $c_1$, and $c_1$, the coefficient of determination $R^2$ can be used as a goodness of fit. The coefficient of determination for the fits (a), (b), (c), and (d) are 0.9999, 0.9998, 0.9999, and 0.9992, respectively.

The total hemispherical emissivity of the selective coating can be estimated given the effective emissivity and the total hemispherical emissivity of glass. Using $\varepsilon_2 = 0.88 \pm 0.02$ for borosilicate glass at 300 K \cite{Jones2019}, $r_1/r_2 = 0.86 \pm 0.02$ for the evacuated tube used in the experiment, and using Eq. \ref{eq:3}, we find the emissivity of the selective coating on the inner tube to be $\varepsilon_1 = 0.072 \pm 0.001$.

The lumped capacitance model used here is valid if resistance to conduction and convection within the fluid is much less than resistance to radiation at the boundary of the fluid, thereby allowing the fluid to be near thermal equilibrium with itself. Convection within the fluid leads to a more constant temperature distribution due to mixing; hence, only conduction within the fluid can be considered as a conservative estimate. The ratio of conductive thermal resistance within the fluid to radiation from the surface of the inner tube is given by the Biot number for radiation, \cite{Incropera2007}
\begin{align}
    Bi_r = \frac{h_r}{kL} = \frac{\varepsilon_{eff} \sigma (T_{1,0} + T_{2,0})(T_{1,0}^2 + T_{2,0}^2) }{kL}, \label{eq:9}
\end{align}
where $h_r = \varepsilon_{eff} \sigma (T_{1,0} + T_{2,0})(T_{1,0}^2 + T_{2,0}^2)$ is the linearized radiative heat transfer coefficient linearized at inner tube temperature $T_{1,0}$ and outer tube temperature $T_{2,0}$, $k$ is the thermal conductivity of the fluid, and $L$ is the characteristic length, given by the ratio of the volume to the surface area of the body (in this case, the inner tube). Using the calculated effective emissivity of the tube $\varepsilon_{eff} = 0.0711$ and taking the linearization about $T_{1,0} = 300$ K and $T_{2,0} = 400$ K, we find that $Bi_r = 0.00879$ for water and $Bi_r = 0.0315$ for ethanol. This implies that conductive transfers are much greater within the fluid than radiative transfers out of the fluid, so the lumped capacitance model for the fluid is appropriate

%%%%%%%%%%%%%%%%%%%%%%%%%%%%%%%%%%%%%%%%%%%%%% Discussion %%%%%%%%%%%%%%%%%%%%%%%%%%%%%%%%%%%%%%%%%%%%%%%%%%
\section{Discussion} \label{sec:discussion}

The derivation of Eq. \ref{eq:6} shows that not all problems relating to radiative heat transfer between two bodies can be treated in the form of Eq. \ref{eq:1}, which is only applicable to an evacuated tube solar collector with coaxial geometry when the transmittance of the glass outer layer is zero. The experiments conducted here show that making the approximation that the transmittance of the outer glass layer is zero is appropriate, given the good fit achieved using the effective emittance equations. This may not always be true however, especially in cases where the inner tube temperature lies outside the range we have studied, for example, where concentrated sunlight is incident on the tube. An interesting point to note about the differences between Eqs. \ref{eq:1} and \ref{eq:6} is that the latter no longer has symmetry in the absolute value of the heat transferred under the exchange of $T_1$ and $T_2$. That is, the magnitude of the heat transferred will depend on which surface is hotter. As a result, the transient behavior of the tube depends on if it is being heated or cooled by radiation transfer between the surfaces. In the practical operating regime for these tubes, which is approximately between the ambient air temperature and the boiling point of water, this is not an issue. However, in cases where $\tau_2 \approx 0$ is no longer a good approximation, the heating and cooling characteristics will no longer be symmetrical. This effect comes about simply as a consequence of the transmitting properties of the outer glass layer and the gray body approximation.

The uncertainty in all the fitting parameters is small, and the RMS deviation is also smaller than 0.1 in all cases. The RMS deviation is smaller and $R^2$ is higher in both heating experiments. The instability of the outer glass temperature due to the temperature controller in the heating experiments will cause such errors, as well as the asymmetry predicted by Eq. \ref{eq:6} not being accounted for in Eq. \ref{eq:8}. Nonetheless, the resulting deviation is still on the order of the accuracy of the temperature sensors, and the coefficient of determination is still acceptably high.

\begin{figure}[H]
    \centering
    \includegraphics[width=0.75\linewidth]{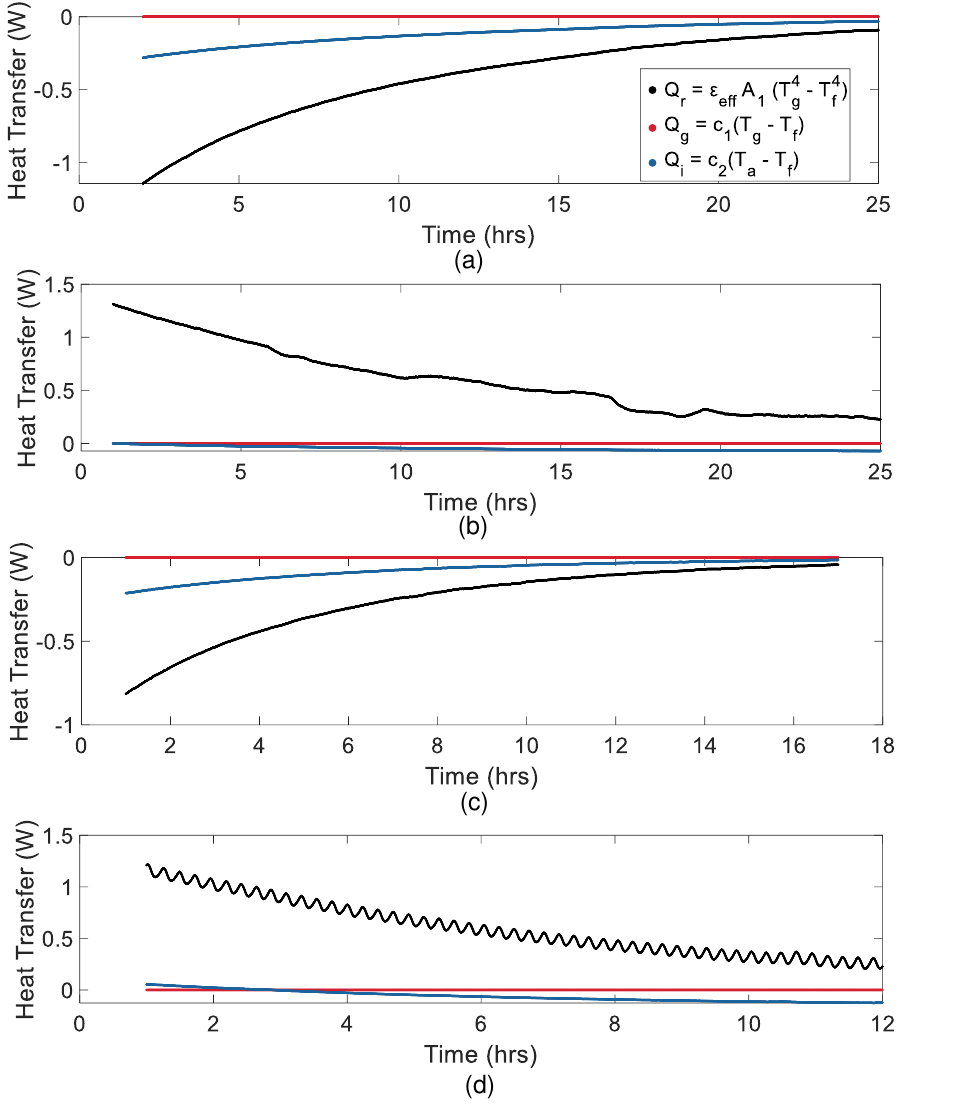}
    \caption{Different heat loss components for each experiment over time. (a) and (b) correspond to the cooling and heating experiments, respectively, with water; (c) and (d) correspond to the cooling and heating experiments, respectively, with ethanol.}
    \label{fig:Fig7}
\end{figure}

From Table \ref{tab:1}, the fitting process found that $c_1 = 0$. Being the coefficient associated with conductive heat transfer through the glass, this result indicates that a negligible amount of heat energy was conducted through the glass. As discussed, this is expected since the conduction through the end of the glass envelope is a high resistance path. Using each term in Eq. \ref{eq:8} and the fitted parameters in Table I, the heat transferred per unit time throughout the experiment for each mode of heat transfer is shown in Fig. \ref{fig:Fig7}. In all cases, especially far away from equilibrium, the total heat transfer is dominated by radiative heat losses. Also in Table \ref{tab:1}, the fit for $c_2$ differs more between the heating cases than is observed between the cooling cases. An explanation of this observation is that, unique to the ethanol heating case, the fluid temperature crosses from being below ambient to being above ambient temperature. The convection coefficient for the heat transfer between the insulation and the ambient air changes as the direction of heat transfer changes. When the ethanol is at a temperature below ambient, heat flows into the tube through the insulating cap, while when it is above ambient, heat flows out. Generally, this would require a different heat transfer coefficient for each case. Here, a single effective heat transfer coefficient is used that still provides a good fit to the model as only a small region of the experiment is below ambient temperature, and the convective heat transfers are small compared to radiative ones.

The fact that $c_1 = 0$ and radiation dominated the heat transfer indicates that very little heat was lost through conduction to the outer envelope due to residual gas. If there was degradation of the vacuum, there would be a larger heat loss term proportional to $(T_g - T_f)$. These results indicate that little vacuum degradation has occurred, and the tube is quite stable. A minor degradation of the vacuum is consistent with the results obtained from the accelerated aging study by Chow et al \cite{Chow1985}.

The exact emissivity of this tube of the Nitto Kohki type mass produced in the 1980s, when new is unknown, but was approximately 0.05 at 100$^\circ$C \cite{Harding1982, Harding1981}. Chow reported an emissivity of $0.050 \pm 0.0006$ for a similar tube before the aging process and an emissivity of $0.065 \pm 0.002$ after aging at 430$^\circ$C for approximately 10\,000 h. These emissivities apply at 100$^\circ$C. The emissivity of the surface calculated in this article was $0.072 \pm 0.001$, which is somewhat higher than the emissivity after accelerated aging but nevertheless demonstrates the ability of these tubes to maintain vacuum and a functional low emissivity coating over long times. The Nitto Kohki tube was fitted with a getter, which may help in removing small amounts of gas.

\begin{figure}[H]
    \centering
    \includegraphics[width=0.98\linewidth]{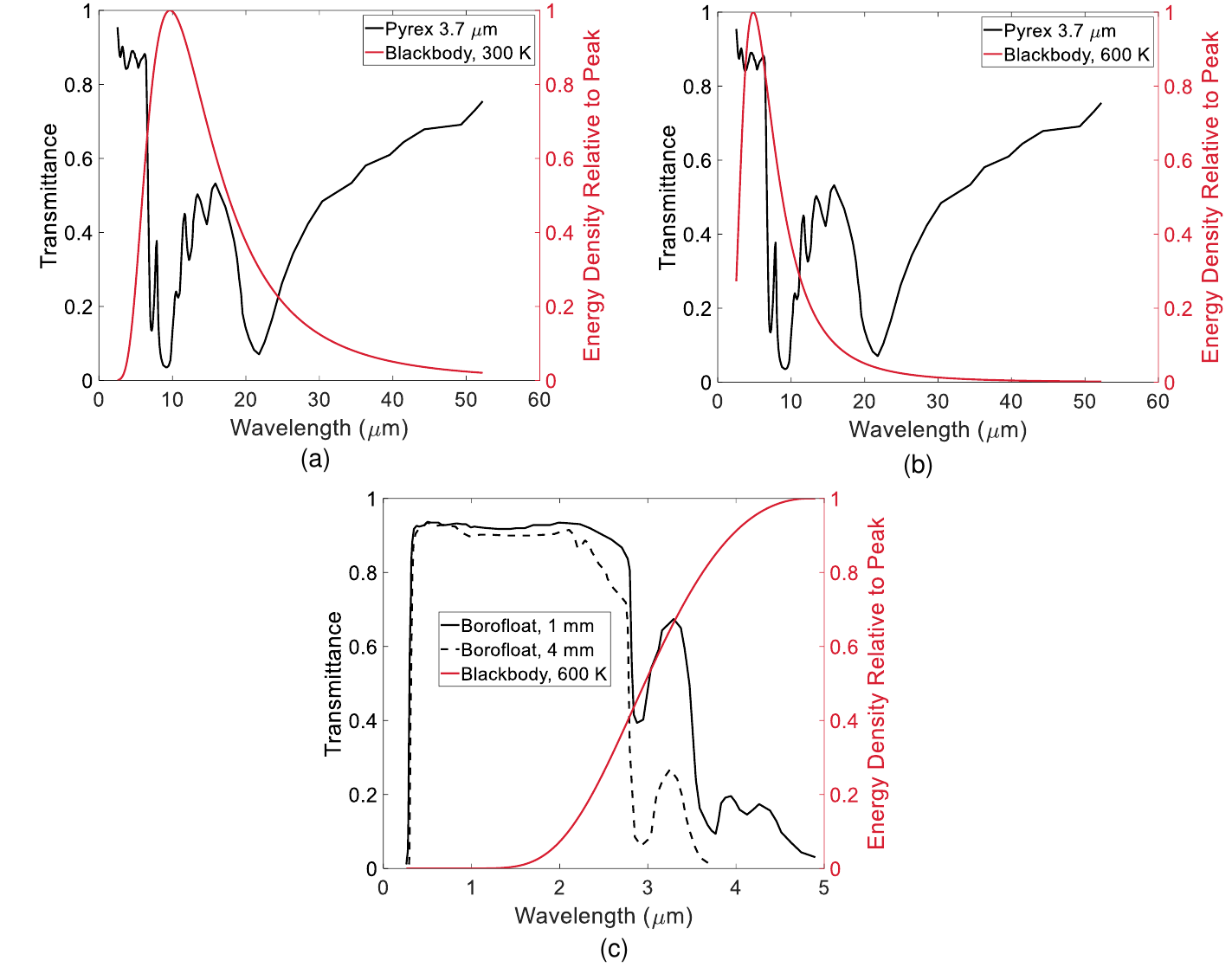}
    \caption{Blackbody distribution superimposed with spectral hemispherical transmittance at 300 (a) and 600 K (b) for 3.7 $\mu$m thick Pyrex samples\cite{Husung1990}, and 600\,K for 1 and 4\,mm Borofloat samples \cite{Hoon2009}.}
    \label{fig:Fig8}
\end{figure}

The above results support the assumption that transmitted radiation through the outer envelope can be neglected for most purposes, that is $\tau_2 = 1 - \varepsilon_2 - \rho_2 \approx 0$ in Eq. \ref{eq:6}. The fraction of blackbody radiation transmitted as a function of its temperature can serve as an upper limit to the heat loss through transparency for inner tubes at various temperatures. Figure \ref{fig:Fig8} shows a normalized blackbody distribution, along with the transmittance of borosilicate glass of thickness 3.7\,$\mu$m, 1\,mm, and 4\,mm \cite{Hoon2009, Husung1990}. The 3.7\,$\mu$m glass samples are thinner than those found in the tubes; however, they serve as a qualitative indicator at higher wavelengths. In the case of the thicker glass, transmittance will be less.

At lower temperatures, near 300\,K, the peak of the blackbody curve lies near minima of the transmittance. As the temperature increases to 600\,K, the peak of the blackbody distribution approaches a transmitting window at the lower end of the near-infrared region, and the total power being radiated is larger. This has practical applications where the temperature of the inner tube is high, as in the delivery of process heat. At higher temperatures, the transmission of the outer envelope means that there is a greater loss of heat from the tube during its operation, accounted for using Eq. \ref{eq:6}, and the assumption that $\tau_2 = 1 - \varepsilon_2 - \rho_2 \approx 0$ will no longer be valid. In this way, the effective transmittance of the outer envelope will have a temperature dependence. Furthermore, the experimental procedure and the model given by Eq. \ref{eq:8} used for determining the effective emittance will no longer be exact, and the definition of an effective emittance is no longer valid.

In the case where the evacuated tube geometry is used to encapsulate a moisture sensitive solar cell, the use of a relatively thin borosilicate outer tube would reduce its operating temperature by radiative transmission through the borosilicate outer tube. The use of a non-reactive gas fill would further increase the heat loss and allow the use of the thinner glass.

From the results in the article, some comments can be made on the practical aspects of the fundamental physics involved in radiation heat transfer in the coaxial geometry. Kirchhoff’s law is stated as the equality of the angular and spectral absorptivities, i.e., $\varepsilon_{\theta, \lambda} = \alpha_{\theta, \lambda}$, and debate has surrounded the equality of the total hemispherical absorptivities and emissivities \cite{Johnson2019, Robitaille2009}. The gray body approximation is a separate consideration. The results have shown that these assumptions in practice provide excellent predictions for the radiative heat transfer heat in the evacuated tubular collector in the temperature regime we have considered.

%%%%%%%%%%%%%%%%%%%%%%%%%%%%%%%%%%%%%%%%%%%% Conclusions %%%%%%%%%%%%%%%%%%%%%%%%%%%%%%%%%%%%%%%%%%%%%%%%%%%
\section{Conclusions}
The evacuated solar collector is an important form of renewable heat generation device and understanding its performance and properties in terms of radiative heat transfer is crucial. Our experimental results showed that the Nitto Kohki tube maintained a stable vacuum after approximately 40 years of storage, and that although the selective coating had undergone minor degradation, the degradation was only slightly higher than obtained in previous accelerated aging studies. We have derived an expression for the radiative heat transfer in this coaxial geometry, showing that the equations need to be modified when the transmission of the outer layer is accounted for. As a result, there is an inherent asymmetry in the cooling/heating of the tube. For low-temperature applications below the boiling point, it is sufficient to neglect the transmissivity of the layer and use the effective emissivity for calculating heat transfer in the nominal operating temperature of the system. The experiment was found to adhere well to the effective emittance approximation. For high-temperature operation, greater than 200$^\circ$C, depending on the wall thickness the effect of the transmitting layer needs to be accounted for using the extended theory presented here since the spectral distribution of the emitted radiation will fall within the transmitting region of the glass.

\newpage
\bibliographystyle{unsrt} 
\bibliography{main}  %%% Remove comment to use the external .bib file (using bibtex).

\end{document}